\documentclass[a4paper,11pt]{article}
\pdfoutput=1

\usepackage{pack} 
\usepackage[T1]{fontenc}

\title{\boldmath Two distinct phases in the first 13 seconds of GRB110731A prompt emission}

\author{Mohammad. A. F. Basha}

\affiliation{Physics Department, Faculty of Science, Cairo University,\\Giza, Egypt}
\affiliation{Principal Investigator of Astrophysics and Space Research Division, Center of Advanced Interdisciplinary Sciences, Faculty of Science, Cairo University,\\Giza, Egypt}

\emailAdd{mafbasha@sci.cu.edu.eg}

\abstract{In this work, the time-resolved BAT/GBM/LAT joint spectral analysis of GRB110731A during the prompt phase from the GBM trigger and up to 13 seconds later showed that, at the very early phase of prompt emission, the emission mechanism is closest to the standard fireball model. This model over-predicts the thermal photospheric emission and used to contradict observations. Lightcurves at different energy bands revealed two distinguishable phases that may come from different regions. First, we have an early phase, which is not detected by LAT, and is dominated by lower energies, which arises from the photospheric emis­sions without any emissions involved in dissipation mech­anisms and characterized by low Lorentz factor and high ra­diation efficiency. This is followed by a later phase, having a more complex structure that remarkably follows the same track in all energy bands and is attributed to emissions from internal shocks. This burst is a good candidate to study both thermal and non-thermal emissions, since the two phases can be clearly separated in lightcurve and spectrum. The rapid variation of Lorentz factor and the values of photo­spheric radii, which are relatively far away from the central engine in Phase 2, are more consistent with the mechanism of collisional heating in baryonic jets. Further information can be obtained by combining more wavelengths with the help of the other detectors.}


\begin{document}
\maketitle
\flushbottom

\section{Introduction}
\label{sec:intro}

\noindent  Gamma-ray bursts (GRBs) are energetic electromagnetic transients that usually consist of two main phases. First, we have an early phase called ``prompt emission'', which lasts for seconds before fading away and has most of the electromagnetic output in the gamma-ray band. This is followed by a later phase, predominated by longer wavelength radiations, called ``afterglow emission'', which can last for a longer time and has most of its electromagnetic output in the X-ray, optical, and radio bands.

\noindent Details of the mechanisms of both emissions are still a matter of debate, but extensive study of GRB spectra from different instruments has led us to strongly believe that the prompt emission is of internal origin, produced in an ultra-relativistic outflow of the internal shocks (Rees and Mészáros 1994; Zhang et al. 2006), while the afterglow is the emission from the forward shock that propagates in the external medium (Mészáros and Rees 1997; Sari et al. 1998).

\noindent Since its launch in November 2004, the Swift satellite (Gehrels et al. 2004, 2005) has helped in studying prompt emission depending on the sophisticated on-board localization capability of its Burst Alert Telescope (BAT) which has a relatively low energy coverage (15--150 keV), with an energy resolution of 5 keV (Barthelmy et al. 2005) along with a large effective area (1000 cm2 at 20 keV in a source onaxis).

\noindent On the other hand, studying emission in the GeV energy range has become possible since the launch of Fermi Gamma-ray Space Telescope on June 2008. The on-board Large Area Telescope (LAT, 20 MeV--$\sim $300 GeV) (Atwood et al. 2009) and Gamma-ray Burst Monitor (GBM, 8 keV-- 40 MeV) (Meegan et al. 2009) together supply important information on the ultra-relativistic outflows of GRBs.

\noindent A joint Fermi/Swift analysis (e.g. Page et al. 2009; Francisco et al. 2011, Stamatikos et al. 2008; Basak and Rao2011) combines the precise localization and low energy response of Swift with the broader spectral coverage of Fermi detectors to study the stages of ultra-relativistic outflows through a wide range of energy bands. This increased the importance of studying GRBs detected by both instruments, which provides more informative and confirmative results regarding spectral parameters and their various correlations.

\noindent This work utilizes BAT/GBM/LAT joint analysis to study the prompt emission of GRB110731A, which was caught by several observatories (Oates et al. 2011, GCN Report 343.1). Its lightcurve is characterized by a complex structure showing multiple-peak pulses. Information regarding the observation and the procedure of data reduction is presented in Sect. ~\ref{sec:research}. The results regarding the characteristics of the time-integrated and time-resolved spectral properties of GRB110731A are presented in Sect. ~\ref{sec:results} and discussed in Sect. ~\ref{sec:discussion}.

\noindent 
\section{Observation and data reduction}
\label{sec:research}

\noindent BAT was triggered (Target ID 458448) at 11:09:37 UT on the 31st July 2011 (Oates et al., GCN Report 343.1). The T${}_{90}$ (15 - 350 keV) for this GRB is 38.8 $\pm$ 13.0 s. The BAT on-board calculated location is RA, 18${}^{h}$42${}^{m}$3${}^{s}$.1 and Dec, -28${}^\circ$32'10", with an uncertainty of 3 arc-minutes. XRT observations and settled UVOT observations began \~{} 56 s and 75 s, respectively, after the BAT trigger. The best position is the UVOT location RA(J2000) = 18${}^{h}$42${}^{m}$0${}^{s}$.99 and Dec(J2000) = -28${}^\circ$32'13 8" with an error of 0.5 arcsec (radius, 90\% confidence). The non-detection in the ultra-violet filters suggests a redshift between 2 and 3, which agrees with the Gemini-N reported redshift z = 2.83 (Tanvir et al., GCN Circ. 12225).

In consistence with the Swift position, the Fermi Gamma-Ray Burst Monitor triggered and located GRB110731A (trigger 333803371 / 110731465) at 11:09:29.94 UT (Gruber, et al., GCN Circ. 12221); it was also detected by Fermi LAT (Bregeon et al. 2011, GCN 12218) showing high energy emission above 100 MeV, with a significance greater than 10 sigma. The angle from the Fermi LAT boresight is 6 degrees, predicting 25-50 counts above 20 MeV in the LAT FOV. Moreover, this burst was bright enough to result in a Fermi spacecraft autonomous rapid repoint (ARR) maneuver.

BAT data were downloaded from the Swift Archive Download Portal\footnote{$ $\url{http://www.swift.ac.uk/swift_portal/}} of the UK Swift Science Data Centre. The data were reduced using the HEAsoft tools (version 6.12), and the latest CALDB files at the time of analysis.

Background-subtracted-mask-weighted lightcurves were extracted using BATBINEVT at the desired energy bands with uniform bin sizes starting from the GBM trigger time.

Both BAT time-integrated and time-resolved spectra were extracted at the default CALDB 80 channels of energy bins, and the systematic error vector was applied using BATPHASYSERR to account for residuals in the response matrix. The BAT detector response matrix is generated using the BATDRMGEN tool which creates a single RSP file, which is then used in the spectral analysis.

Fermi GBM data were downloaded from the online Fermi GRB burst catalog on the NASA HEASARC web­site\footnote{$ $\url{http://heasarc.gsfc.nasa.gov/W3Browse/fermi/fermigbrst.html}} and LAT data were obtained using FSSC's web site data server\footnote{$ $\url{http://fermi.gsfc.nasa.gov/cgi-bin/ssc/LAT/LATDataQuery.cgi}} (Fermi SSC-LAT Photon, Event and Space­craft Data). Background-subtracted time-dependent spectra and energy-dependent lightcurves for GBM and LAT were constructed following Zhang et al. (2011).

Fermi Science Tools (v9r27p1) and the HEASOFT tools (version 6.11.1) are used to extract uniformly binned lightcurves from the Time Tagged Event (TTE) files of the NaI triggered detectors n${}_{0}$ and n${}_{3}$ and the BGO detector b${}_{0}$ for the different energy bands. PHA2 files containing spectra for the desired time cuts are extracted where the background is calculated from 40 to 20 seconds of pre-burst data.

The same bin sizes were also used to extract LAT lightcurves for energy bands above 100 MeV which are found to show one main peak with an emission duration of about 13 seconds. For the purpose of this work, the prompt emission is investigated through a simultaneous analysis of different energy bands within this time interval, particularly, from T${}_{0}$ -- 0.5 to T${}_{0}$ + 13 seconds.

The LAT spectra from 100 MeV to 200 GeV with corresponding response files were constructed for the desired time cuts using the ``P7TRANSIENT\_V6'' class. For this class the relevant timescales are sufficiently short and the additional residual charged particle backgrounds are much less significant. Also, the short LAT emission duration makes the accumulation of LAT background counts negligible. To justify the significance of GRB110731A, a likelihood analysis is applied using the standard Science Tools software package provided by the FSSC. The analysis confirmed detection with a TS value of 159 and a number of background predicted counts (N${}_{pred}$) of   3.62 x 10${}^{-4}$ from an isotropic component along with a number of 6.87 x 10${}^{-6}$ counts for the standard galactic background component with N${}_{pred}$ = 6.87 x 10${}^{-6}$ indicating a very small number of background contributions from both components. Also an estimation of the count rate using the pre-burst data, which were re-scaled to the 13 s time range, showed that the background is not more than just one photon in this time range.

Analysis of BAT/GBM/LAT joint spectra was performed using the X-Ray Spectral Fitting Package (Xspec v.12.7.1) (Arnaud 1996). The Cash/Castor statistics method is not suitable for BAT spectra which are created already having been background-subtracted and having non-Poissonian errors (Page et al.  2009). In this case, the chi-square method is the appropriate method for these data. However, since a likelihood-based statistic is essential for the analysis of the low-count LAT data, the PGSTAT option in XSPEC (Arnaud 1996), which allows for applying different statistics to the same group of data, is used to provide reliable errors. In this way the likelihood statistics as implemented in this option can be applied to the data with a Poissonian nature.

For joint fits, normalizations of all GBM detectors were tied together and the cross-calibration followed the procedure in Page et al.  2009, by first allowing GBM normalization to vary while keeping the BAT and LAT normalizations frozen at unity and then the average relative constant is used to link the normalizations of the three detectors which led to significantly improve the statistics.

\begin{table}[tbp]
\tiny
\centering
\begin{tabular}{|c|c|c|c|c|c|c|c|c|c|}
\hline

Time from T${}_{0}$  & \multicolumn{5}{|c|}{Band + PL} & \multicolumn{4}{|c|}{CPL + PL} \\ \hline 
(s) & ${\rm A}$ & $\beta $ & E${}_{p}$ (keV) & $\Gamma _{PL}$ & $\chi $${}^{2}$/dof & $\Gamma _{CPL}$ & E${}_{p}$ (keV) & $\Gamma _{PL}$ & $\chi $${}^{2}$/dof \\ \hline 
-0.5 -- 13 & -- 0.53 $\pm$ 0.14 & -- 2.95 $\pm$ 0.56 & 244 $\pm$ 45 & 1.77 $\pm$ 0.07 & 491/423 & -- 0.48 $\pm$ 0.14 & 241 $\pm$ 37 & 1.74 $\pm$ 0.05 & 489/424 \\ \hline 
1 -- 11 & -- 0.14 $\pm$ 0.16 & -- 3.26 $\pm$ 0.70 & 192 $\pm$ 29 & 1.75 $\pm$ 0.06  & 503/423 & -- 0.16 $\pm$ 0.15 & 243 $\pm$ 29 & 1.71 $\pm$ 0.04 & 504/424 \\ \hline 
2 -- 9 & -- 0.21 $\pm$ 0.14 & $<$ -- 3.5 & 279 $\pm$ 33 & 1.75 $\pm$ 0.05 & 530/423 & + 0.038 $\pm$ 0.15 & 259 $\pm$ 26 & 1.74 $\pm$ 0.04 & 528/424 \\ \hline 
3 -- 7 & -- 0.01 $\pm$ 0.17 & -- 2.97 $\pm$ 0.45 & 264 $\pm$ 29 & 1.93 $\pm$ 0.13 & 543/423 & + 0.080 $\pm$ 0.14 & 298 $\pm$ 28 & 1.83 $\pm$ 0.08 & 551/424 \\ \hline 
4 - 5 & -- 0.49 $\pm$ 0.19 & $<$ -- 5 & 330 $\pm$ 72 & 1.99 $\pm$ 0.01 & 440/423 & + 0.155 $\pm$ 0.23 & 282 $\pm$ 34 & 1.88 $\pm$ 0.22 & 434/424 \\ \hline 

\end{tabular}
\caption{\label{tab:a} The results of fitting the time intervals under study for the GBM/BAT data using Band+PL and CPL+PL models. }
\end{table}

\section{Results}
\label{sec:results}
 
\noindent Figure 1A represents the GBM and LAT lightcurves at different energy bands using uniform bin size of 1 second. The figure shows the two phases to be discussed, an early phase, which is not detected by LAT and dominated by energies lower than 100MeV, followed by a later phase, having a more complex structure that remarkably follows the same track in all energy bands.

\noindent Signal to noise (S/N) ratio trials from 5 and up to 30 are applied to lightcurves from the BAT detector, making use of its sensitivity to inspect the high variability of the different peaks in the lightcurve complex structure. After that, these peaks are followed in both NaI and BGO lightcurves and the rising phase of each peak was combined in one bin. These bins are represented by the seven different slices in Figure 1B, where the counts are re-binned using a uniform bin size of 0.5 s to provide a good visual representation of the intensity behavior.

\begin{figure}[tbp]
\centering 
\includegraphics[width=.9\textwidth]{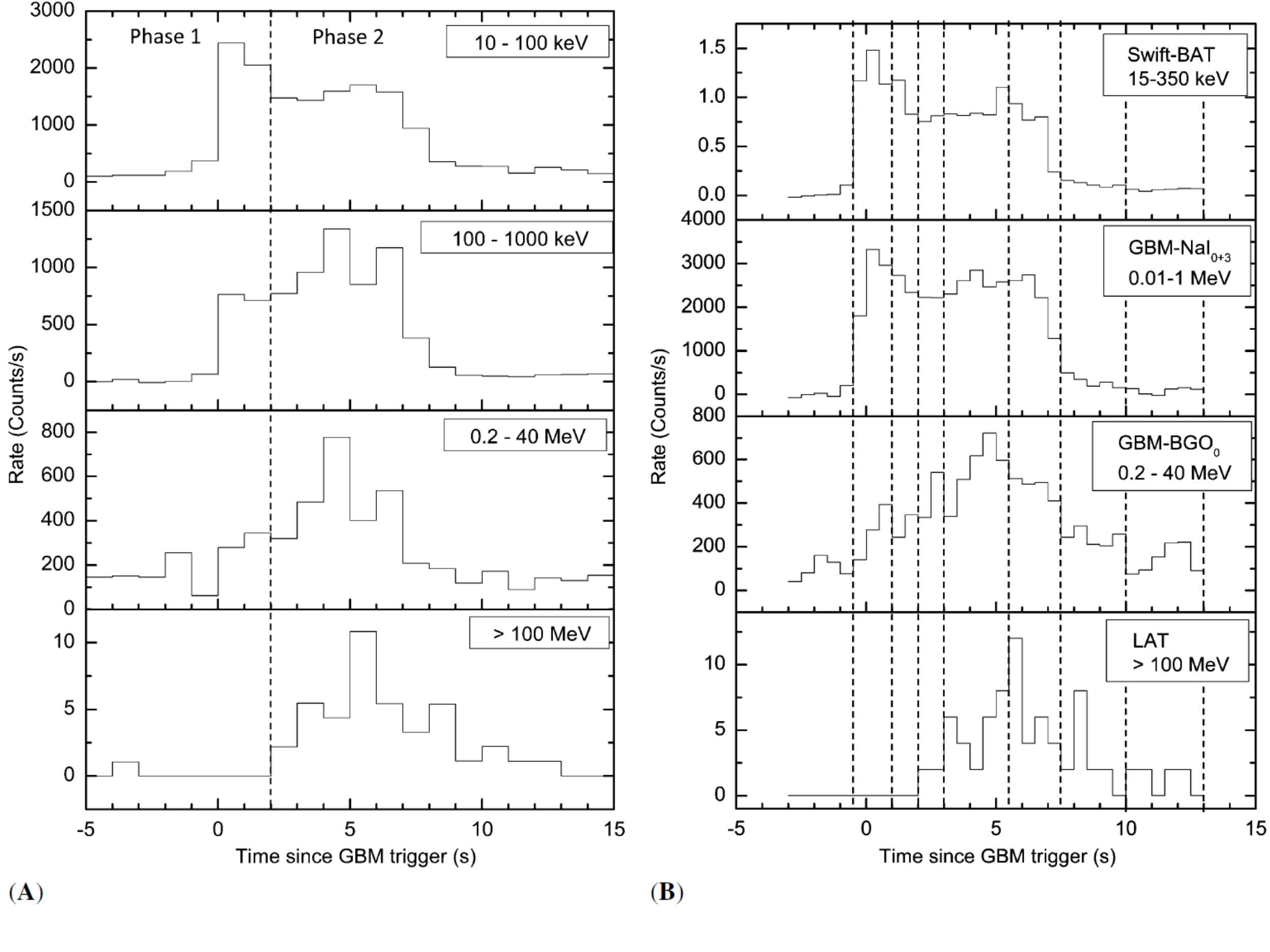}
\caption{\label{fig:a} (A) GBM and LAT lightcurves at different energy bands using a uniform bin size of 1 second, where the vertical dashed line separates the two phases under study. (B) The uniformly binned lightcurves, with bin size 0.5 s, for the different detectors, where the vertical dashed lines indicate the intervals under study.}
\end{figure}

Spectra of each detector are simultaneously constructed at these intervals, noting that no emissions are detected in the LAT band during the first and second time slices.

The time-integrated spectrum is the result of the spectral evolution over time. It is typically described in the literature by the Band model or the cutoff-power law model. Attempts have been made to explain the physical interpretation of the Band shape of the spectra with the help of the synchrotron emission hypothesis. This succeeded in explaining many aspects of the emission, although it failed as regards the unpredicted hard-low-energy spectra of some bursts (Ghirlanda et al. 2003, Crider et al. 1997). On the other hand, the photospheric emission hypothesis, which assumes thermal-blackbody radiation, was able to account for the hard-low-energy spectra of these bursts (M´esz´aros et al. 2002, M´esz´aros \& Rees 2000). The use of black-body models contributes in explaining the physics of GRBs by providing parameters with physical significance.

The broadening in the spectra of most bursts leads to the result that a single narrow black-body component is not adequate. In some cases, the time-integrated spectrum is a superposition of different black-bodies with time-dependent temperatures and fluxes.

The origin of the GRB110731A prompt emission was investigated here by first attempting to fit the time-integrated joint BAT/GBM spectrum using models of non-thermal origin. There is agreement that the best fit model is always the one that has the smallest number of parameters, unless adding other components improves the chi-squared value by 6 for each additional parameter (Sakamoto et al. 2007).

A series of time-resolved spectral analysis are carried out here similar to the analysis in Zhang et al\textit{.}~2011 on GRB090902B to investigate the time smearing effect on the spectrum. The Band function (Band et al. 1993) showed an acceptable fit, which improved after the addition of a simple power-law (PL) component ($\chi ^{2}$/dof = 491/423) giving a peak energy E${}_{p}$ = 166 $\pm$ 24 keV, a low energy spectral index $\Gamma $ = --0.53 $\pm$ 0.14 and high energy spectral index $\Gamma $ = --2.95 $\pm$ 0.56 indicating a narrow curvature Band component. Zooming into the time interval T${}_{0}$ -- 1 to T${}_{0}$ -- 11s resulted in Band spectral indices describing a spectrum of narrower curvature ( $\Gamma $ = --0.14 $\pm$ 0.16,  $\Gamma $ = --3.26 $\pm$ 0.70, and E${}_{p}$ = 130.03 $\pm$ 16.76 keV). As the time bin gets smaller, the curvature of the Band component becomes progressively narrower.

\begin{figure}[tbp]
\centering 
\includegraphics[width=.9\textwidth]{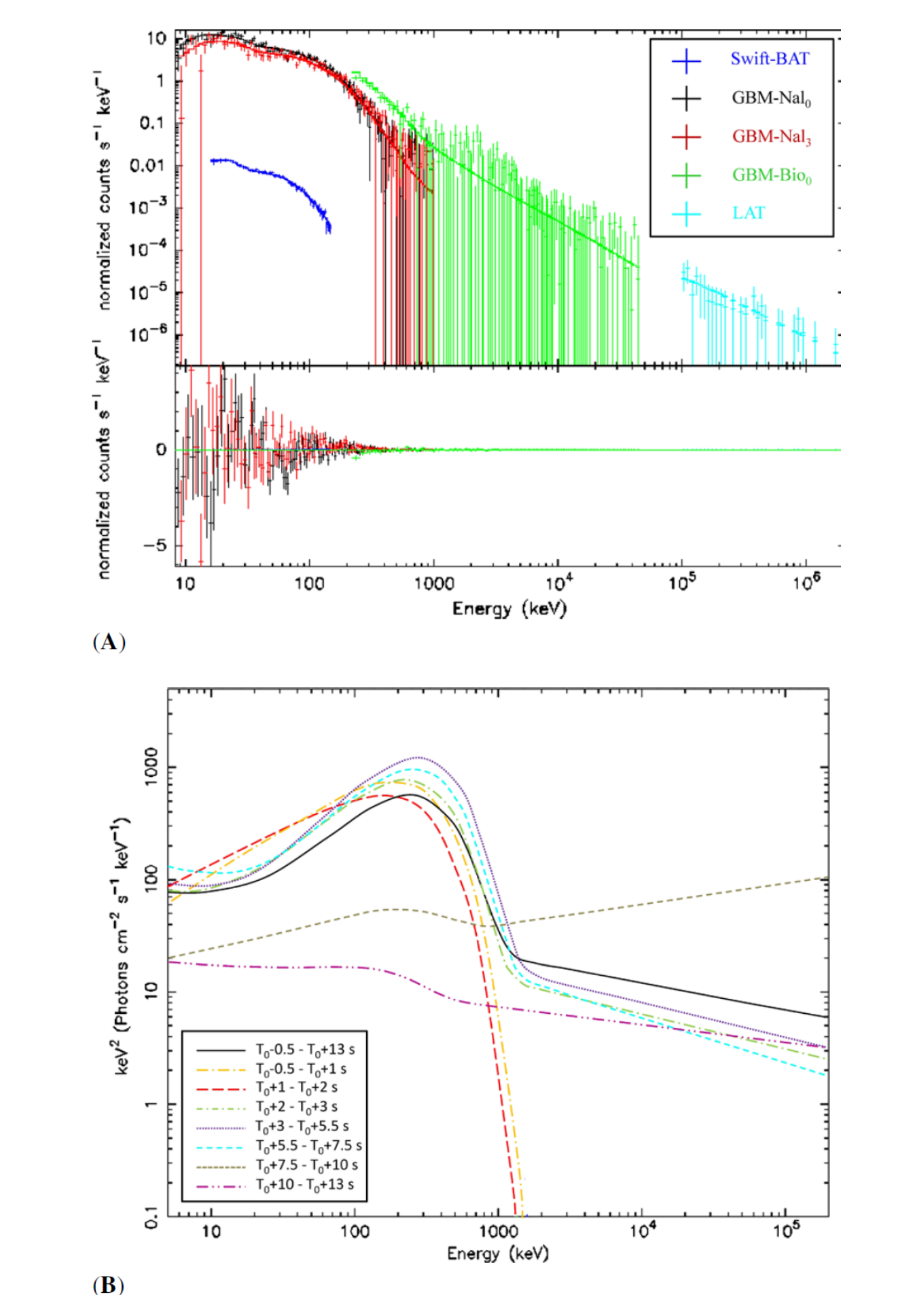}
\caption{\label{fig:b}  (A) The fitting of time-integrated joint BAT/GBM/LAT spectrum (top panel) by mBB function representing the thermal component plus the PL function representing the non-thermal component, where the residuals are plotted in terms of normalized counts/sec/keV (bottom panel). (B) The spectral evolution of$ \upsilon F_{ \upsilon }$ through the seven intervals under study of the time-resolved joint BAT/GBM/LAT spectra.}
\end{figure}

The Cut-off Power-Law (CPL) model with additional PL component succeeded in fitting the spectrum as well ($\chi ^{2}$/dof = 489/424), with photon index value of $ \Gamma  _{CPL}$ = --0.48 $\pm$ 0.14. This decreases gradually when zooming into the same time intervals as before. At the smallest time interval the CPL+PL fitting returned a reasonable fit ($ \chi ^{2}$/dof = 448/425) when fixing  $\Gamma _{CPL}$ = +1, which is the Rayleigh-Jeans slope of a black-body (BB). Table 1 shows the results of fitting the indicated time intervals by Band+PL and CPL+PL models.

The fact that the observed spectra are a superimposition of narrower Band components along with the behavior of CPL model at the low energy regime indicates a possible contribution from a black-body component. Thus one may suspect that the Band-like spectrum of GRB110731A prompt emission could be either a result of a superposition of thermal and non-thermal components (Ryde 2004, Ryde \& Pe'er 2009) or it could be a temporal superposition of many black-body-like components.

To investigate these possibilities, the spectrum is fitted with a model in which the peak of the emission is provided by a single Planck function. To account for any non-thermal emission, an additional non-thermal component is included and represented by a PL model. This model did not improve over Band+PL and CPL+PL fittings ($ \chi ^{2}$/dof = 540/425).

Considering the possibility that the broadening of the spectrum may be attributed to contributions from emissions arising from different regions in space, a deviation from a simple Planck function is expected, which is represented by a broadened photospheric component on the form of a multicolor black-body (mBB) model (Ryde et al 2010 and Larsson et al. 2011). This model provided an acceptable fit with $ \chi ^{2}$/dof = 513/426; however, adding a PL function to allow for an additional non-thermal component has significantly improved the quality of the fitting, resulting in $ \chi ^{2}$/dof = 488/424.

Recent Fermi observations have revealed additional power-law components at high energies in a number of bright GRBs (Abdo et al. 2009, Ackermann et al. 2010). Since GRB110731A was detected by LAT, a PL function is expected to improve the fit, however, without the LAT data, the contribution from the PL component to the total model is effectively small and the photon indices could not be constrained properly in some time intervals Thus, introducing the LAT data here helped to add more adequate counts to constrain the non-thermal component for all time intervals. However, since this work focuses more on the thermal component and no LAT data were detected for the first two intervals; all fluxes presented here are calculated over the energy range from 10 keV to 45000 keV for comparison between all intervals.

The time-integrated joint spectrum for all detectors is illustrated in Figure 2 A, where the residuals are plotted in terms of normalized counts/sec/keV. The same model is also used to fit the seven intervals of the time-resolved joint BAT/GBM/LAT spectra. Due to low LAT counts, the PGSTAT option is used here instead of the chi-square method as mentioned before. Adding the PL component to the spectra of the first two intervals did not improve the quality of the fitting or affect the shape of the curve, indicating that this component is not needed in such intervals. The spectral evolution of the best fit $ \upsilon F_{ \upsilon }$ spectra of all time bins is demonstrated in Figure 2 B and the resulting parameters are presented in Table 2, where the statistical result STAT is the sum of C-stat + chi-square.

The evolution of both the fraction of the flux due to the thermal component (F${}_{BB}$/F${}_{tot}$) and the temperature (kT) as a function of time are shown in Figure 3. During the first and second time slices, no emissions are detected by LAT, and the thermal flux dominates the initial emission phase up to $\sim $ 2 s after GBM trigger. At later times, the non-thermal flux clearly takes over, as seen in Figure 3 A.

Figure 3 B indicates a distinct break at \~{} 5 s at which \textit{kT} \~{} 85 keV. Using the value of the black-body temperature at the break and assuming a standard cosmology with $\Omega $${}_{\Lambda }$ = 0.73, $\Omega $${}_{M}$ = 0.27 and H${}_{0}$ = 71 km s${}^{-1}$ Mpc${}^{-1}$, I have followed Pe'er et al. 2007 to estimate the coasting value of the outflow Lorentz factor ($\Gamma $ \~{} 765 Y${}_{0}$${}^{1/4}$) and the initial radius during expansion (r${}_{0}$ \~{} 3.46 x 10${}^{8}$ Y${}_{0}$${}^{-3/2}$) Here Y${}_{0}$ is the ratio of the fireball energy and the energy emitted in gamma rays. The photospheric radius (r${}_{ph}$), beyond which the outflow becomes optically thin, is found to be \~{} 1.29 x 10${}^{12}$ Y${}_{0}$${}^{1/4}$ cm. This value is larger than the saturation radius (r${}_{s}$ \~{} 2.65 x 10${}^{11}$ Y${}_{0}$${}^{-5/4}$ cm). The parameters r${}_{ph}$, r${}_{0}$, and r${}_{s}$, calculated for different time intervals, are presented in Table 2, where the term (r${}_{ph}$ /r${}_{s}$)${}^{-2/3}$ indicates the thermal efficiency (Beloborodov 2010).

\begin{figure}[tbp]
\centering 
\includegraphics[width=.9\textwidth]{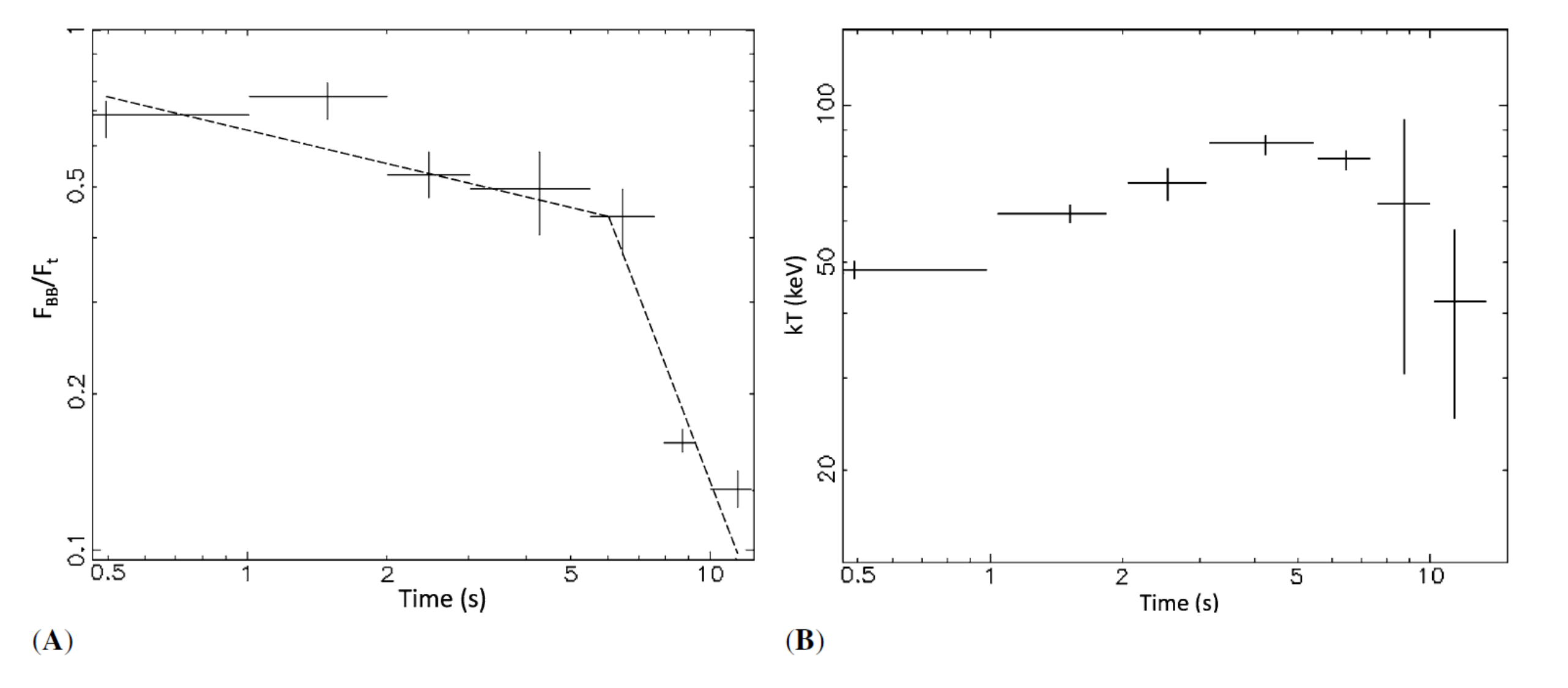}
\caption{\label{fig:c} (A) The evolution of the fraction of the flux due to thermal component (F${}_{BB}$/F${}_{tot}$) with time, indicating the domination of the thermal flux through the initial emission phase up to $\sim $ 2 s after GBM trigger. (B) The evolution of the temperature kT with time indicating a distinct break at \~{} 5 s at which \textit{kT} \~{} 85 keV. }
\end{figure}

The parameters of the outflow as a function of time are illustrated in Figure 4. Since the bulk Lorentz factor $\Gamma $  is most strongly dependent on the temperature, the evolution of these quantities is expected to track each other (Ryde et al. 2010). As seen in Figure 4, $\Gamma $ starts to rise from a value of $\Gamma _{0}$ \~{} 489 Y${}_{0}$${}^{1/4}$, to its maximum value of $\Gamma _{m}$ \~{} 746 Y${}_{0}$${}^{1/4}$ which is at the break time of the temperature decay and then decreases afterwards. The time-averaged value for the bulk Lorentz factor is $\Gamma _{av}$ \~{} 594.6 Y${}_{0}$${}^{1/4}$, with a standard deviation of 115.5.

\begin{figure}[tbp]
\centering 
\includegraphics[width=.9\textwidth]{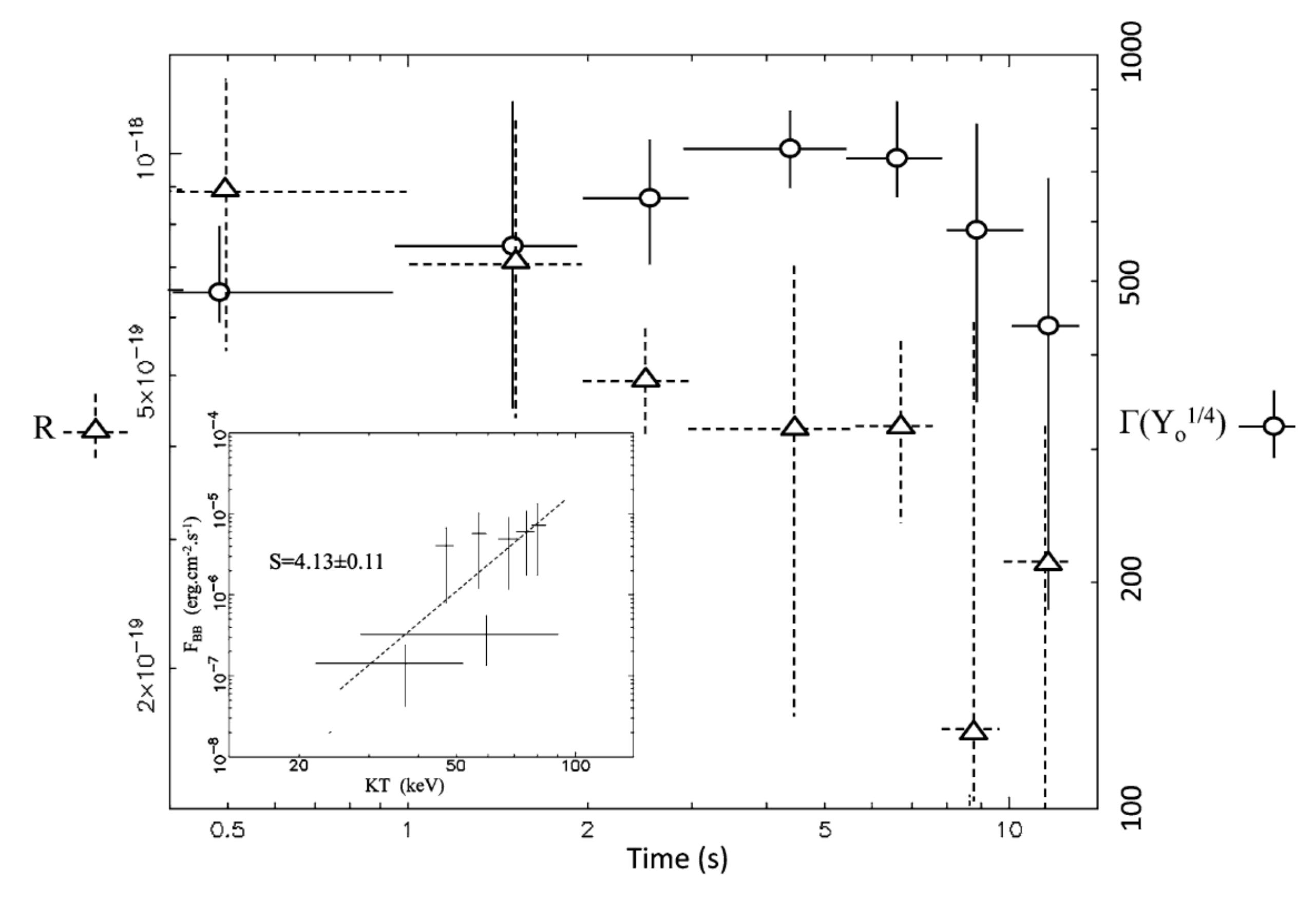}
\caption{\label{fig:d} The ratio between the observed flux and the emergent flux, R = ($F_{BB}/ \sigma T^{4}$)${}^{1/2}$, as a function of time (left axis) and the change of the bulk Lorentz factor ($\Gamma$) with time (right axis). The inset represents the correlation between the black-body flux (F${}_{BB}$) and the temperature (kT) showing the fundamental property of a black-body emitter in which the observed flux F${}_{BB}$ is proportional to T${}^{4}$. }
\end{figure}

The time evolution of the parameter R = (F${}_{BB}/\sigma$ T${}^{4}$)${}^{1/2}$, which describes the ratio between observed flux and the emergent flux for individual time slices, varies randomly with time within the same order of magnitude as seen in Figure 4. The typical R value of most bursts varies by an order of magnitude, as reported in Ryde \& Pe'er 2009, and this indicates that the behavior of R here deviates from the universal behavior of increasing during the rising phase of each pulse. Since GRB110731A has a complex inseparable structure, the apparent constancy of \textit{R} with time may be attributed to interference of \textit{R} values from different pulses. Ryde \& Pe'er 2009 have shown that some bursts with complex structure are consistent indeed with having constant \textit{R} throughout the analyzed time period. This behavior goes with the interpretation of R as an effective transverse size of the emitting region. The plots of the black-body flux versus temperature for these bursts show that the observed flux F${}_{BB}$ $\propto $ T${}^{4}$ which is the fundamental property of a black-body emitter. Comparing this to the inset of Figure 4, which represents the hardness-intensity correlation for GRB110731A, one finds that the observed black-body flux (F${}_{BB}$) increases with the fourth power of the temperature (kT), and is therefore proportional to the emergent flux, $\sigma T^{4}$, consistent with the previous discussion.

\section{Discussion}
\label{sec:discussion}

\noindent The time-resolved spectral analysis revealed two phases in the lightcurve with different spectral properties. These different characteristics, which are exhibited simultaneously in the prompt emission of one single GRB, appear to indicate that there might be different origins in the different prompt emission episodes of some GRBs (Zhang, Fu-Wen 2012).

\begin{figure}[tbp]
\centering 
\includegraphics[width=.9\textwidth]{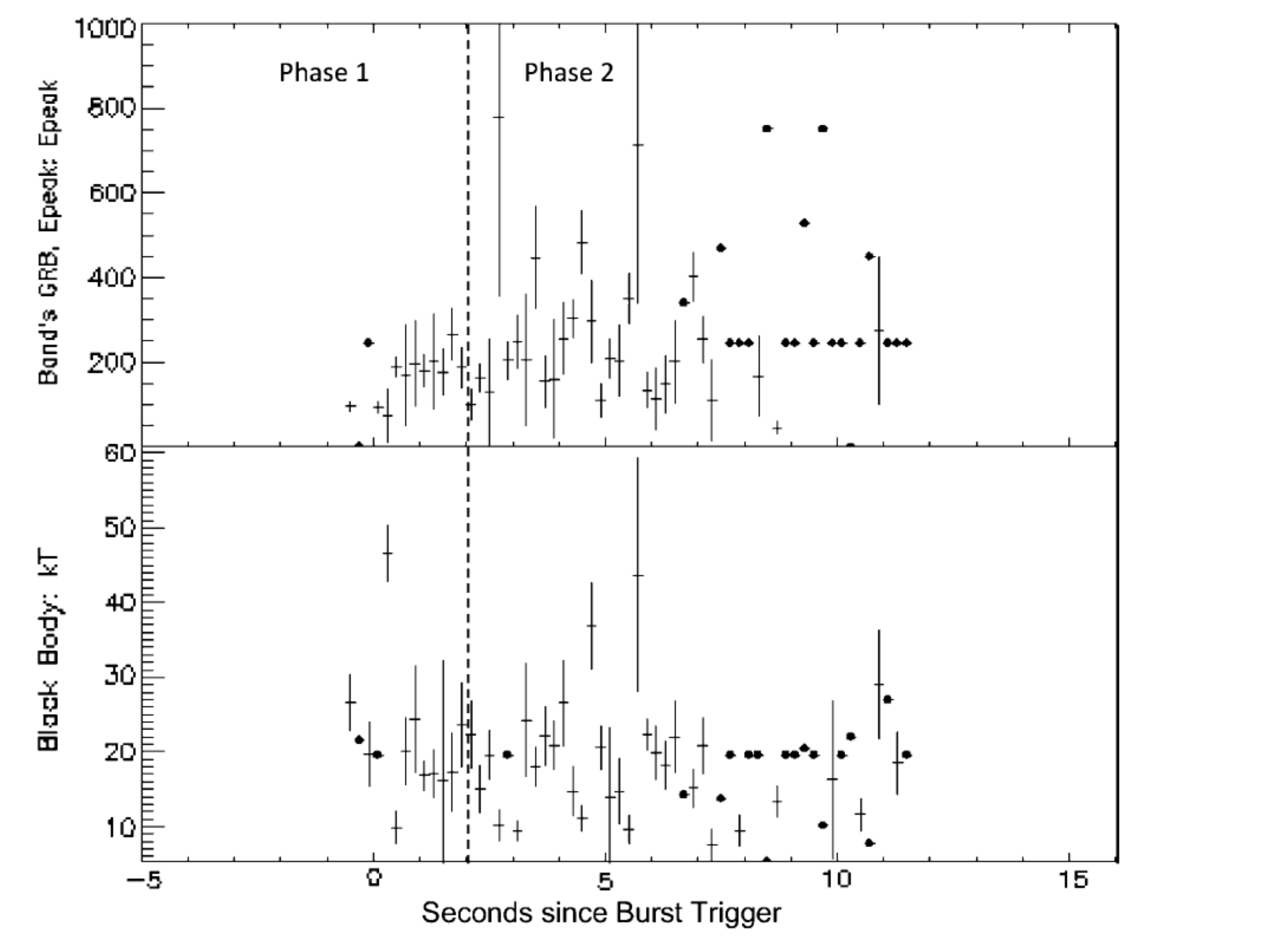}
\caption{\label{fig:e}  The evolution of the Band function E${}_{peak}$ (top panel) and the BB temperature kT (bottom panel) over the duration of the burst showing a quite stable temperature that has a weak correlation with E${}_{peak}$. The vertical dashed line separates the two phases under study. }
\end{figure}

To confirm whether the two thermal and non-thermal components here, are due to two different mechanisms, a study of the temporal development of each component is performed here which is similar to the analysis by Guiriec et al. 2010 on GRB 100724. These authors divided the lightcurve of the GBM data into a number of intervals to follow the evolution of the Band function E${}_{peak}$ and the black-body (BB) temperature kT over the duration of the burst. Here the GBM-only data were analyzed using the Fermi RMFIT package (V3.3) developed by the GBM Team for GBM and LAT analysis. To remove the need for the BAT data for this analysis, I have added to the GBM data two more detectors with a good geometry to the source (NaI${}_{6}$ and BGO${}_{0}$) to improve the statistics of the GBM-only analysis. In terms of the Castor C-STAT values, the Band + BB fits has better statistics (C-STAT/dof = 800/620) than the Band only fit, with E${}_{peak}$ = 246 $\pm$ 16.7 keV and kT = 19.6 $\pm$ 1.58 keV. The brightness of the burst helped in following the evolution of the spectral components through 58 time intervals. This shows a quite stable temperature that has a weak correlation with E${}_{peak}$, as seen from Figure 5. The kT distribution, which is depicted in Figure 6, can be fit by a Gaussian function and has a mean value of 20.83 $\pm$ 1.9 with a standard deviation of 5.66 $\pm$ 2.1. This is consistent with the temperature obtained from the time-integrated spectral fit, suggesting that the BB component does not evolve much over time while E${}_{peak}$ of the non-thermal component follows the typical variations over time.

\begin{figure}[tbp]
\centering 
\includegraphics[width=.9\textwidth]{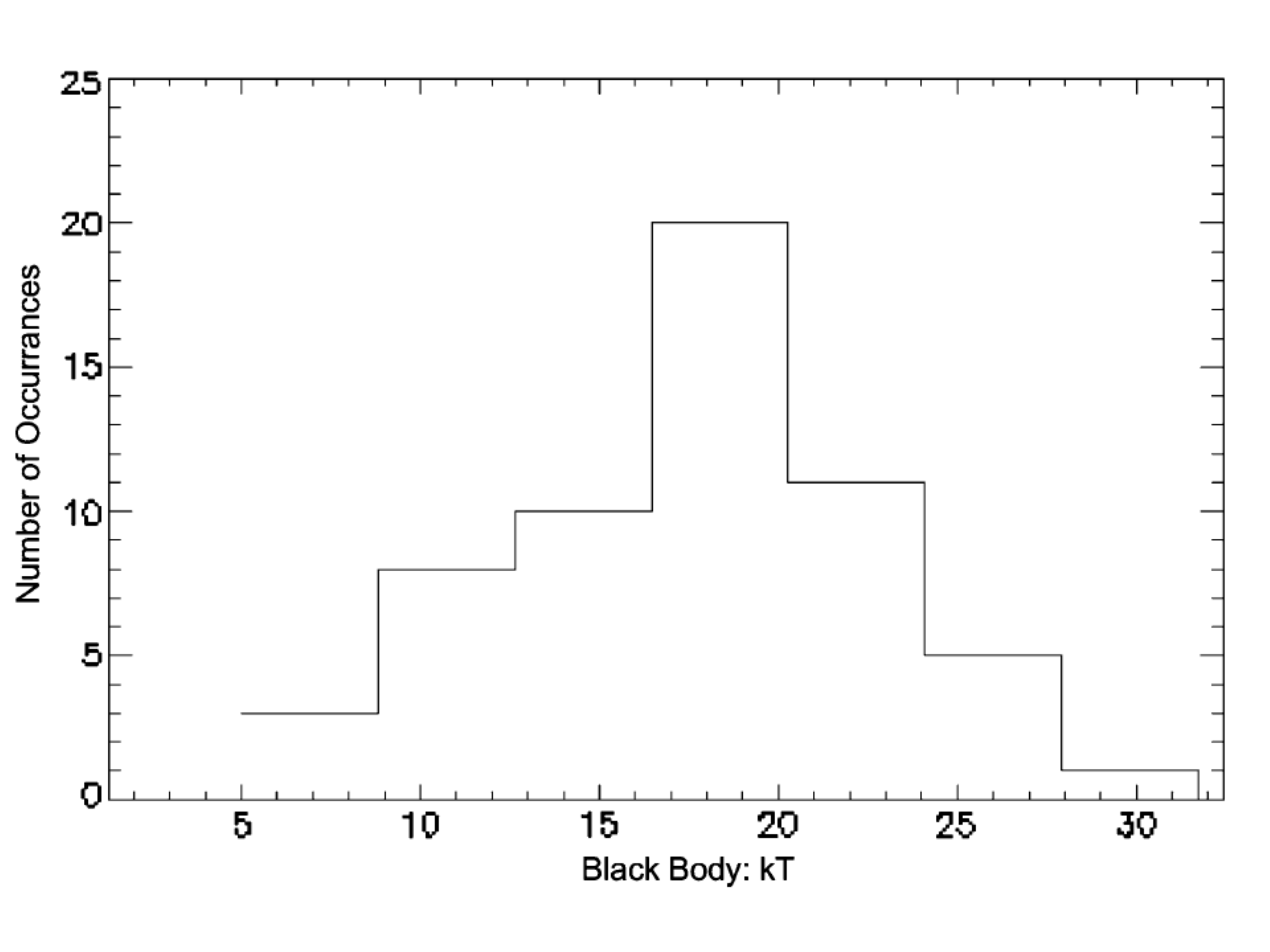}
\caption{\label{fig:f}  Distribution of the time-resolved black-body temperature kT. The most probable occurrence is consistent with the temperature obtained in the time-integrated spectral fit. }
\end{figure}

The fact that these two components seem to vary independently favors the idea that the presence of the BB in the time-integrated spectrum cannot be attributed to spectral evolution of the Band function during the burst. This means that this thermal component is of photospheric origin, while the non-thermal emission occurs at larger radii. The expected photospheric emission in GRB spectra was early suggested on such theoretical grounds by Goodman 1986, M´esz´aros 2002 and Rees \& M´esz´aros 2005, among others.

As the results indicated, Phase 1 (Figure 1A) is associated with pure photospheric emission and does not involve any dissipation processes. It is characterized by high thermal efficiency and a maximum fraction thermal flux equal to 74.5\% of the total flux. This phase is not detected by LAT causing an onset delay which is a common feature of most LAT-detected bursts (e.g., Abdo A. A. et al. 2009 on 080916C, Abdo A. A. et al. 2009 on 080825C, Ackermann M. et al. on 090510, Abdo A. A. et al. 2009 on 090902B).

The delayed onset of the $>$100MeV emission from the GBM trigger has been modeled for GRB 080916C as arising from proton synchrotron radiation in the prompt phase (Razzaque et al. 2009) and for GRB 090510 as arising from electron synchrotron radiation in the early afterglow phase (Kumar \& Barniol Duran 2009; Ghirlanda et al. 2009). It was often difficult for the photospheric model to explain the delayed onset of the $>$100MeV emission.

\begin{table}[tbp]
\tiny
\centering
\begin{tabular}{|c|c|c|c|c|c|c|c|c|c|}
\hline

Time from  & kT${}_{max}$& $q=4-\frac{2}{p} $ & PL index & STAT/dof & E${}_{p}$${}^{*}$& r${}_{o}$ x 10${}^{8}$ & r${}_{s}$ x 10${}^{11}$ & r${}_{ph}$ x 10${}^{12}$ & (r${}_{ph}$ /r${}_{s}$)${}^{-2/3}$ \\

T${}_{0}$ (s) & (keV) & & &  & (keV) & (Yo${}^{-3/2}$ cm) & (Y${}_{o}$${}^{-5/4}$ cm) & (Y${}_{o}$${}^{1/4}$ cm) & (Y${}_{o}$${}^{-1}$) \\ \hline

-0.5 -- 1 & 75.72 $\pm$ 2.2 & 0.68 $\pm$ 0.052 & 2.3 $\pm$ 0.24 & 463/424 & 127 $\pm$ 11.8 & 15.5 $\pm$ 7.1 & 7.57 $\pm$ 1.21 & 2.07 $\pm$ 0.47 & 0.51 $\pm$ 0.08 \\ \hline 
1 -- 2 & 89.38 $\pm$ 2.3 & 0.81 $\pm$ 0.062 & 2.5 $\pm$ 0.35 & 457/424 & 145 $\pm$ 26.8 & 13.7 $\pm$ 4.7 & 7.52 $\pm$ 0.84 & 1.85 $\pm$ 0.25 & 0.55 $\pm$ 0.05 \\ \hline 
2 -- 3 & 95.26 $\pm$ 4.7 & 1.28 $\pm$ 0.039 & 2.3 $\pm$ 0.57 & 380/333 & 248 $\pm$ 16.3 & 5.69 $\pm$ 0.3 & 3.64 $\pm$ 0.43 & 1.52 $\pm$ 0.16 & 0.39 $\pm$ 0.05 \\ \hline 
3 - 5.5 & 106.92 $\pm$ 3.1 & 1.56 $\pm$ 0.019 & 2.3 $\pm$ 0.30 & 489/437 & 325 $\pm$ 17.4 & 4.38 $\pm$ 0.8 & 3.27 $\pm$ 1.17 & 1.55 $\pm$ 0.19 & 0.35 $\pm$ 0.04 \\ \hline 
5.5 - 7.5 & 100.54 $\pm$ 4.2 & 1.48 $\pm$ 0.018 & 2.4 $\pm$0.46 & 487/436 & 316 $\pm$ 15.2 & 3.73 $\pm$ 1.1 & 2.70 $\pm$ 0.22 & 1.51 $\pm$ 0.22 & 0.32 $\pm$ 0.03 \\ \hline 
7.5 - 10 & 87.83 $\pm$ 17 & 0.48 $\pm$ 0.10 & 1.8 $\pm$ 0.29 & 491/432 & 194 $\pm$ 7.10 & 0.30 $\pm$ 0.07 & 0.18 $\pm$ 0.07 & 0.52 $\pm$ 0.51 & 0.1 $\pm$ 0.03 \\ \hline 
10 -- 13 & 60.8 $\pm$ 9.5 & 0.58 $\pm$ 0.39 & 2.1 $\pm$ 1.1 & 415/429 & 57 $\pm$ 24.30 & 0.37 $\pm$ 0.12 & 0.16 $\pm$ 0.03 & 1.09 $\pm$ 0.10 & 0.06 $\pm$ 0.01 \\ \hline 
\textbf{-0.5 - 13} & \textbf{102.7 $\pm$ 3.9} & \textbf{1.36 $\pm$ 0.14} & \textbf{2.2 $\pm$ 0.17} & \textbf{497/451} & \textbf{289 $\pm$ 8.6} & \textbf{3.46 $\pm$ 1.12} & \textbf{2.65 $\pm$ 0.67} & \textbf{1.29 $\pm$ 0.33} & \textbf{0.35} $\pm$ 0.05\textbf{} \\ \hline

\end{tabular}
\caption{\label{tab:b} The fitting parameters and the parameters of the outflow from the time-integrated and time-resolved BAT/GBM/LAT spectra. (${}^{*}$ E${}_{p}$ is determined by a cutoff power-law fit) }
\end{table}

The missing high energy emission can be explained here according to the fact that the early phase of GRB is pure photospheric emission dominated by energies quite far below 100 MeV. The duration of this phase is very small on the time scale of the source rest frame (\~{} 0.5 s); however, this duration differs from one GRB to another.

Emissions from internal shocks and small magnetic dissipation start to take over in Phase 2, during which the lightcurves of GBM and LAT track each other roughly in all energy bands up to 7.5 seconds. After that, the fireball starts to cool down. The time-resolved spectral analysis of Phase 2 indicated a comparable contribution from two components, a thermal component due to kinetic dissipation, which is modeled by a multicolor black-body and a non-thermal component, which is adequately described by a simple power law and strongly dominates the emission after 7.5 seconds.

The relatively high values of the bulk Lorentz factor ($>$ 700) during Phase 2 suggest a bright photospheric emission, as predicted by the standard theory (Paczy´nski 1990), usually with a quasi-thermal spectrum (Ryde 2005). However, the strong thermal component dominating the emission during Phase 1, when the coasting bulk Lorentz factor is less than 600, may imply that one should exclud magnetically dominated outflows, this predicts a smeared thermal component.

The rapid variation of Lorentz factor and the values of photospheric radii that are relatively farther away from the central engine (Table 2) are more consistent with the collisional heating mechanism in a baryon loaded jet (Beloborodov 2010), where a hot $e^{\pm}$ plasma is created as a result from the nuclear and Coulomb collisions in the GRB jet. The standard model of a baryonic jet assumes comparable numbers of neutrons and protons. 

The high energy spectral index $\Gamma  \sim $ 2.95 obtained from fitting the time-integrated joint spectrum with Band+PL function suggests radiations from two competing Comptonization mechanisms. One is thermal, which dominates the early times of Phase 2. This is due to $e^{\pm}$ that are continuously thermalized by Coulomb collision with protons inevitably heated to a relativistic temperature by nuclear collisions. The other mechanism is non-thermal due to irradiative cooling of $e^{\pm}$ injected by inelastic nuclear collisions. The contributions from thermal and non-thermal emissions are comparable during Phase 2 up to 7.5 seconds, which is consistent with the standard model of baryonic jets (Beloborodov 2010), while the domination of the non-thermal emission, at later times during Phase 2, could be due to synchrotron emission or inverse Compton emission from dissipation regions outside the photosphere (Ryde 2005). However, the steepening of the PL photon index at later times indicates that non-thermal electrons are cooled quickly by the thermal radiation and emit non-thermal Compton radiation. An increase in thermal emission and energy density from the photosphere would lead to an increase in the Compton cooling and emission from the non-thermal electrons. Further information can be obtained by combining the results from other wavelengths with the help of other detectors or from studying the afterglow emission.

\section{Conclusion}
\label{sec:concl}

\noindent Results in the literature (Guiriec et al. 2010) have shown that the most extreme version of the magnetized outflow scenario is not possible, where the energy is released by the central engine as a pure Poynting flux. The results here, however, show that pure photospheric emission is possible in the early phase near the trigger time. The calculated radii at the base of the flow agree with those predicted by the standard fireball model without requiring a very high efficiency for the non-thermal process. The missing high energy emission can be explained here according to the fact that this phase, which has a duration of \~{} 0.5 s on the time scale of the source rest frame, is pure photospheric emission, which is dominated by energies quite far below 100 MeV. The connection between the LAT onset delay and the fact that the early phase of GRB110731A is pure photospheric could be investigated for other bursts. LAT emission appears only when internal shocks and small magnetic dissipation start to take over in Phase 2.

\pagebreak

\end{document}